\documentclass[11pt,a4paper]{article}
\usepackage[utf8]{inputenc}
\usepackage[T1]{fontenc}
\usepackage{lmodern}
\usepackage[margin=2.5cm]{geometry}
\usepackage{amsmath,amssymb}
\usepackage{graphicx}
\usepackage{booktabs,multirow,array,longtable}
\usepackage[table]{xcolor}
\usepackage{setspace}
\usepackage[font=small,labelfont=bf]{caption}
\usepackage[hidelinks]{hyperref}
\usepackage{xurl}

\graphicspath{{figs/}}
\hypersetup{
  pdftitle={Mapping rockfall hazard and detecting precursory damage in rock slopes with passive seismic: Lessons from the La Praz case-study},
  pdfauthor={P. Bottelin, O. Méric, L. Baillet, D. Beniamine, A. Lescurier},
  pdfsubject={Preprint},
  pdfkeywords={Passive seismic, HVSR, Resonance frequency, Unstable rock slope, Early warning system, Rockfall}
}

\newcommand{\aH}{a_{HVSR}(f_p)}
\newcommand{\fp}{f_p}
\definecolor{sdpA}{gray}{0.30}
\definecolor{sdpB}{gray}{0.78}
\definecolor{sdpC}{gray}{0.90}
\definecolor{sdpD}{RGB}{170,20,40}
\newcommand{\Pone}{\cellcolor{sdpA}\color{white}P1}
\newcommand{\Ptwo}{\cellcolor{sdpB}P2}
\newcommand{\Pthree}{\cellcolor{sdpD}\color{white}P3}

\begin{document}

\begin{center}
\fbox{\parbox{0.92\textwidth}{\small\setstretch{1.1}
\textbf{Preprint.} This is the authors' submitted version of the article published in
\emph{Engineering Geology}, 338 (2024), 107627.
The final version is available at \url{https://doi.org/10.1016/j.enggeo.2024.107627}.
This version may differ from the published version.}}
\end{center}

\vspace{1em}
\begin{center}
{\Large\bfseries Mapping rockfall hazard and detecting precursory damage in rock slopes with passive seismic: Lessons from the La Praz case-study\par}
\vspace{1.2em}
{\large Pierre Bottelin$^{a,*}$, Ombeline Méric$^{a}$, Laurent Baillet$^{b}$, David Beniamine$^{c}$, Anne Lescurier$^{d}$\par}
\vspace{0.8em}
{\small\setstretch{1.1}
$^{a}$Société Alpine de Géotechnique SAGE, 2 rue de la Condamine, 38610 Gières, France\\
$^{b}$Centre National de la Recherche Scientifique (CNRS), Institut des Sciences de la Terre (ISTerre), Université Grenoble Alpes, 38000 Grenoble, France\\
$^{c}$Tétras Libre, 8 rue Mayencin, 38410 Saint Martin d'Hères, France\\
$^{d}$Conseil Départemental de la Savoie, Service Risques Naturels, L'Adret, 1 rue des Cévennes, 73000 Chambéry, France\\
$^{*}$Corresponding author: p.bottelin@sage-ingenierie.com\par}
\end{center}

\section*{Abstract}
We conducted repeated successive passive seismic surveys on the La Praz unstable rock slope ($\sim$13,000--15,000~m$^3$, Savoie, France), spanning over more than ten years. The last survey was conducted two months before its complete collapse in August 2023. We used the amplitude of Horizontal to Vertical Spectral Ratio at peak ($\aH$) to discriminate stable from unstable areas and delineate the fractures with major mechanical decoupling effect. We also classified the compartments in terms of rockfall hazard: the most prone-to-fall areas showed the highest $\aH$ (in the range [26--157]), compared to [7.5--19] on intermediate areas and 2.7 on adjacent stable rock mass.

The frequency of HVSR main peak ($\fp$) on three unstable rock compartments showed unambiguous $-15$\%, $-14$\% and $-11$\% drops prior to the collapse. These drops significantly exceed the range of environmentally driven resonance frequency wandering observed on site and generally reported as a limitation in the literature. These drops were interpreted as progressive in-depth slope damage. Interestingly, $\fp$ losses were detected solely on the three most prone-to-fall rock compartments showing the highest $\aH$. The unveiling of in-depth damage processes on such a complex, heterogeneous and highly fractured rock slope using passive seismic is a first.

\medskip
\noindent\textbf{Keywords:} Passive Seismic; Horizontal to Vertical Spectral Ratio HVSR; Resonance Frequency; Unstable Rock Slope; Early Warning System; Rockfall

\section{Introduction}

Unstable rock slopes are widespread within regions with steep topography such as mountains and coastal areas. Risk mitigation may be particularly challenging in populated areas hosting infrastructures, housing and/or people put at stake without vulnerability reduction solutions (Farvacque et al., 2019; Maheshwari et al., 2023). In addition, rock collapse hazard may be relatively sudden as large, detectable and unequivocal creep-like ground displacement pattern is not systematically observed prior to failure (Arosio et al., 2009; Rose and Hungr, 2007; Rosser et al., 2007). Ambient vibration surveys have been increasingly used to detect, characterise and sometimes monitor gravitational hazards (Jaboyedoff et al., 2019), notably because they yield in-depth information. Applications have covered a wide range of unstable volumes ranging from unstable rock compartments, towers and arches (small to moderate volumes i.e. thousands to tens of thousands cubic meters) up to massive rockslides (hundreds of thousands to millions cubic meters), see reviews in Colombero et al. (2021b); Kleinbrod et al. (2019) and e.g. surveys in Finnegan et al. (2021); Moore et al. (2019, 2016); Starr et al. (2015).

As unstable rock sites are generally intensely fractured, they show i) lower bulk stiffness than adjacent stable slope and ii) mechanically independent rock compartments, which can both explain the seismic resonance observed within the unstable body (Burjánek et al., 2010, 2012; Kleinbrod et al., 2017b). Polarization effects and amplification have been systematically reported in the passive seismic records conducted within unstable areas (Burjánek et al., 2014; Colombero et al., 2017; Lotti et al., 2018; Moore et al., 2011). A large set of dynamic parameters (DP) may be computed from ambient vibration surveys, notably resonance frequencies, spectral amplification ratios (e.g. Horizontal to Vertical Spectral Ratio HVSR, Horizontal to Horizontal Spectral Ratio HHSR, Site to Reference Spectral Ratio SRSR; e.g. Bottelin et al., 2021; Burjánek et al., 2019; Colombero et al., 2021a; Häusler et al., 2021b; Kleinbrod et al., 2019, 2017b; Valentin et al., 2017), modal shapes and damping (Brincker and Ventura, 2015; Fu and He, 2001; Häusler et al., 2021a). Benefiting notably from seismic arrays (i.e. the simultaneous operation and time-referencing of sensors on the field), advanced signal processing techniques such as time-frequency polarization analysis (TFPA), frequency-wavenumber (f,k) analysis, enhanced frequency domain decomposition (EFDD) or seismic noise interferometry (dV/V) have been applied recently (Bessette-Kirton et al., 2022; Colombero et al., 2021a; Finnegan et al., 2022, 2021; Geimer et al., 2020; Häusler et al., 2021b, 2021a; Xie et al., 2023).

Passive seismic surveys have been used to delineate the unstable body from adjacent stable slope and to map the degree of instability within rockslides. Such surveys yielding in-depth mechanical information revealed especially useful to delineate hidden fracture nets, using amplification ratios or modal vector amplitude measured at site's resonance frequencies (Burjánek et al., 2018; Häusler et al., 2019; Kleinbrod et al., 2019, 2017a, 2017b). The dynamic parameters (DP) of an unstable slope can also be monitored over time (Geimer et al., 2022; Häusler et al., 2022, 2021b; Starr et al., 2015). This can be achieved through successive punctual surveys or continuous records, the latter being more challenging to maintain in the long term in mountainous areas. The fundamental resonance frequency $f_0$, i.e. the frequency at which the first ---and generally greatest--- amplification peak occurs is generally the clearest DP. Thus, it appears well suited for monitoring purposes (Colombero et al., 2021b). Mono-channel spectra analysis can allow detecting and tracking $f_0$, although spectral amplitude wandering over several orders of magnitude may complicate the monitoring (Bottelin et al., 2021). In contrast, spectral ratio computation (i.e. HVSR or HHSR) offers a very good experimental stability with only little amplitude wandering over time. Spectral ratios also allow easy and robust comparison between data acquired with different instruments at different times. HVSR requires only one sensor at a time on the field, i.e. it does not require the synchronous operation of a reference sensor. However, rock slope resonance does not systematically occur strictly within the horizontal plane in mountainous areas with steep topography (Müller and Burjánek, 2023). Dip was found to be generally inclined downwards the slope for most of the 11 volume-controlled sites studied in Kleinbrod et al. (2019), with a mean dip of about $-5$ to $-10$\textdegree. The spectral ratio amplitude was found very sensitive to dip orientation effects. This may hide resonance modes and/or show peaks that do not coincide with site's resonance frequencies, notably when strong vertical motion occurs. HVSR should hence be used cautiously (Bottelin et al., 2021; Valentin et al., 2017).

In the existing literature, a first study revealed a $-30$\% drop in $f_0$ a few weeks prior to a limestone rock column collapse (Lévy et al., 2010). To this day, this is regrettably the sole case study reporting unequivocal changes in DPs related to rock slope evolution towards failure. Ambient vibration polarization changes that could be related to changes in rock slope's stress field were reported in Häusler et al. (2021b), but no concomitant evidence of slope damage was observed. In contrast, the presence of reversible, environmentally-driven, seasonal or daily changes in DPs has been revealed and discussed in many studies (Bottelin et al., 2013b; Colombero et al., 2021b; Geimer et al., 2022; Häusler et al., 2021b; Larose et al., 2015). Resonance frequencies but also damping ratios and relative velocity changes dV/V have shown to fluctuate with time (Bièvre et al., 2018; Colombero et al., 2018; Häusler et al., 2021b; Le Breton et al., 2021). This wandering appears to be controlled by environmental factors, the predominant factor on most rock slopes being the air temperature (Colombero et al., 2021b). Other parameters may play a role, such as site geometry, fracture network pattern, freeze-thaw cycles, failure mechanism, presence of water, altitude and solar exposition. The amplitude of these reversible changes varies greatly from one site to another, ranging generally from a few percents to tens of percents (Colombero et al., 2021b; Geimer et al., 2022; Kleinbrod et al., 2019). To our knowledge, the largest reversible change in fundamental resonance frequency was reported at Les Arches 1000~m$^3$ unstable column by Bottelin et al. (2013b). An approximate +300\% rise in $f_0$ occurred every winter: the unstable column was being stuck to the adjacent stable slope by a huge amount of ice crystallizing into the main rear crack. Although they indeed reflect the mechanical state of the monitored slope, these environmental fluctuations in DPs may interfere and even hide irreversible changes related to progressive failure processes. Another study focused on monitoring rock column's resonance frequencies during rock bolting works (Bottelin et al., 2017). Increases in resonance frequencies associated with steel rebars and grouting reinforcement were clearly detected, with +7\% in fundamental resonance frequency $f_0$ and up to +17\% in higher mode $f_3$ resonance frequency. This state of the art yet suffers from major limitations. Only two observations of significant and irreversible changes in DPs on unstable rock slopes have been reported (Bottelin et al., 2017; Lévy et al., 2010). Both studies concerned limestone rock columns with clear fractures separating them from adjacent stable massif. Consequently, passive seismic monitoring performance remains to be proven for other rock geology, failure mechanism and geometry. This represents a key element for future integration into practical early warning systems.

In this article, we present a passive seismic study conducted on the La Praz unstable slope (Savoie, French Alps). The $\sim$15,000~m$^3$ unstable volume is made of highly degraded and fractured rock material, with unclear geometry and hidden fracture network. It shows a rather ductile evolution over time with a progressive failure process spanning over many years. First, we investigate the seismic resonance of the site using the HVSR computed on ambient vibrations. Then, the HVSR peak amplitude is used to map the degree of instability across the site. After that, we investigate the temporal changes in site's resonance, benefiting from successive surveys spanning over about ten years.

\section{Site description}
\label{sec:site}

\begin{figure}[p]
\centering
\includegraphics[height=0.6\textheight]{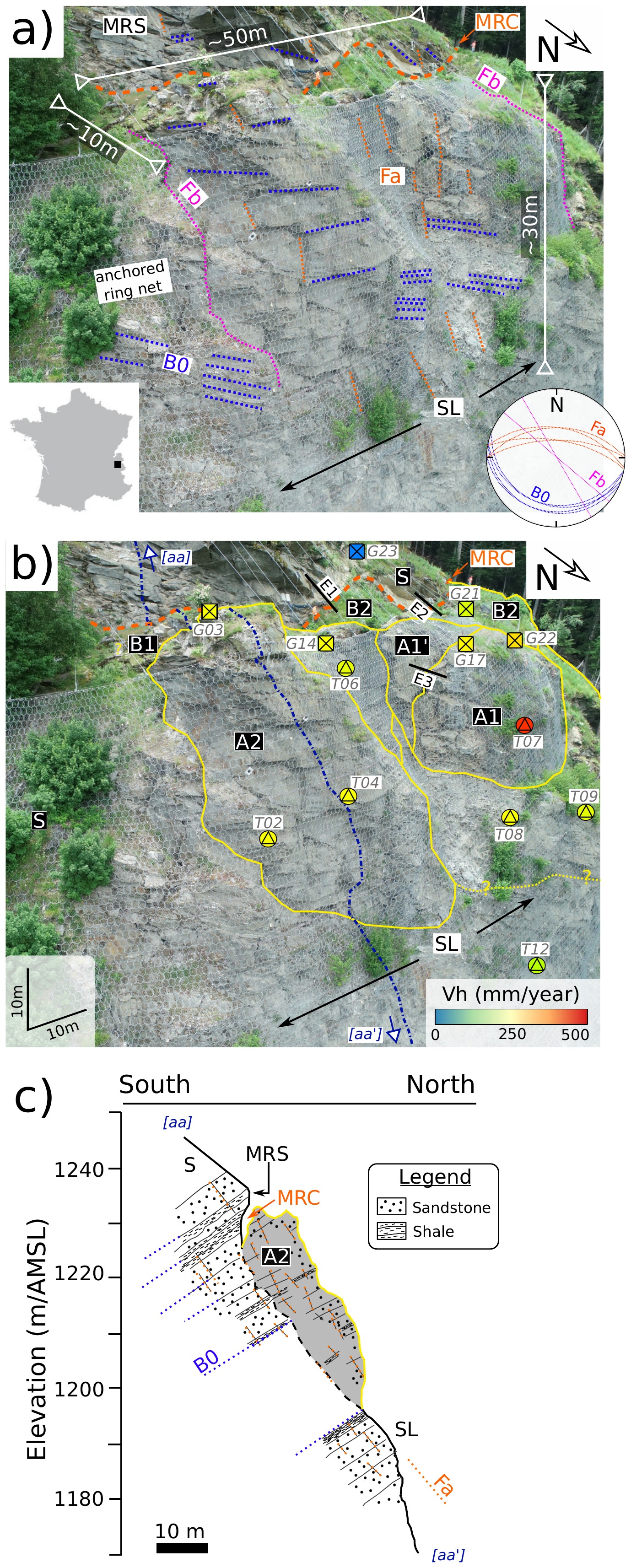}
\caption{a) Location (bottom left inset) and geological description of the site. The North direction is given by the arrow in the top right corner. The figure and the stereo plot (bottom right, Wulff net, lower hemisphere) both show the bedding (B0, blue dotted lines) as well as fracture sets Fa and Fb (orange and purple dotted lines, respectively). The main rear crack (dashed orange line) and main rear scarp bound the site at south. The stable ledge (SL, double headed arrow) is visible within the northern cliff. The unstable slope is about 30~m thick, 50~m wide and 10~m deep for a total volume of $\sim$15,000~m$^3$. b) Unstable slope division into compartments (A1 to S, white label on black background) observed in geotechnical surveys. The outline of each compartment is shown with a yellow solid line. The blue dot-dash line [aa-aa'] corresponds to the vertical cross-section in (c). The Leica total station targets (circles with inscribed triangle, Ti) and the GNSS measurement points (squares with inscribed cross, Gi) are both coloured according to the magnitude of horizontal ground displacement rate (MHGDR) between 25/06/2021 and 21/06/2023 (see colorbar at bottom right). Ti and Gi are tagged with grey italic labels. The extensometers E1, E2, E3 are shown with thick black lines. c) Geological cross-section of the site along cut line [aa-aa'] (see panel b). The unstable compartment A2 is highlighted with grey. Bedding B0 and fracture set Fa are shown as in (a), whereas fracture set Fb is not visible on this section. The geology is shown in the legend.}
\label{fig:site}
\end{figure}

The La Praz site is located in the Maurienne valley (Savoie, French Alps). The unstable area lies at about 1230~m in altitude on the northern-facing flank of the valley (Fig.~\ref{fig:site}a). It is made up of ``Houiller'' series, showing an alternation of heavily fractured sandstone beds and dark mica-rich shale layers of Carboniferous age (Debelmas, 1989). The rock bedding dips into the slope with a 20--30\textdegree S orientation (discontinuity set B0 shown with dotted blue lines in Fig.~\ref{fig:site}a,c). In addition to bedding, the slope is cut by several major fracture sets. Fracture set Fa (orange dotted line) is oriented 090--100\textdegree/55--70\textdegree N. It is nearly collinear with the general slope inclination. Another ---less visible--- fracture set Fb oriented N130--150\textdegree/N90 bounds the eastern and western sides of the unstable slope (purple dotted lines, Fig.~\ref{fig:site}a). The rockslide is about 50~m wide, 30~m high and 10~m thick, for a total unstable volume of about 15,000~m$^3$ (Bottelin et al., 2013a). The combination of rock bedding B0 with the two fracture sets Fa and Fb creates a heavily fractured, unstructured volume of rock blocks, reaching up to 10~m$^3$ unitary volumes for sandstones (Fig.~\ref{fig:site}a,c). The kinematics of the site is complex: it combines rock pile toppling, bed bending, and subsidence at head with possible sliding along a Fa basal surface. A similar complex kinematics has been described e.g. in Gschwind et al. (2019). The progressive subsidence and sliding of the unstable area along the main rear crack exposed a $\sim$10~m high main rear scarp at rear (Fig.~\ref{fig:site}a,c). A more compact rock bed acting as a Stable Ledge (SL) was detected at about 1190--1195~m in altitude during visual inspection on rope and confirmed with active seismic tomography (Bottelin et al., 2013a).

Repeated geotechnical surveys conducted by the SAGE consulting company (not available in public literature) on behalf of the road operator CD73 delineated several homogeneous areas within the unstable mass (compartments A1 to B2, Fig.~\ref{fig:site}b). Compartments A1 and A2 ($\sim$2000~m$^3$ each) located at the edge of the cliff were considered as the most prone-to-fall. Compartment A2 shows a peculiar geometrical shape, forming a slender rock column at its eastern extremity. Compartment A1' ($\sim$1000~m$^3$) is located at rear of A1 and is bounded by two sets of widely opened, deep cracks. The top sides of B1 and B2 compartments are wedged between the main rear crack at south and the other compartments (A1-A1'-A2) at north, while their toes extend towards the cliff face (Fig.~\ref{fig:site}b).

The La Praz site raises major risk mitigation issues because of its location at a narrow gully in the valley. At the slope's toe lie i) the France-Italy 2-ways railway in a covered gallery, ii) a local road belonging to the CD73 operator, iii) a buried France-Italy electrical power line, iv) the Arc river and v) the France-Italy A43 2 $\times$ 2-ways motorway.

\section{Rockfall activity}
\label{sec:rockfall}

\begin{figure}[htbp]
\centering
\includegraphics[width=\textwidth]{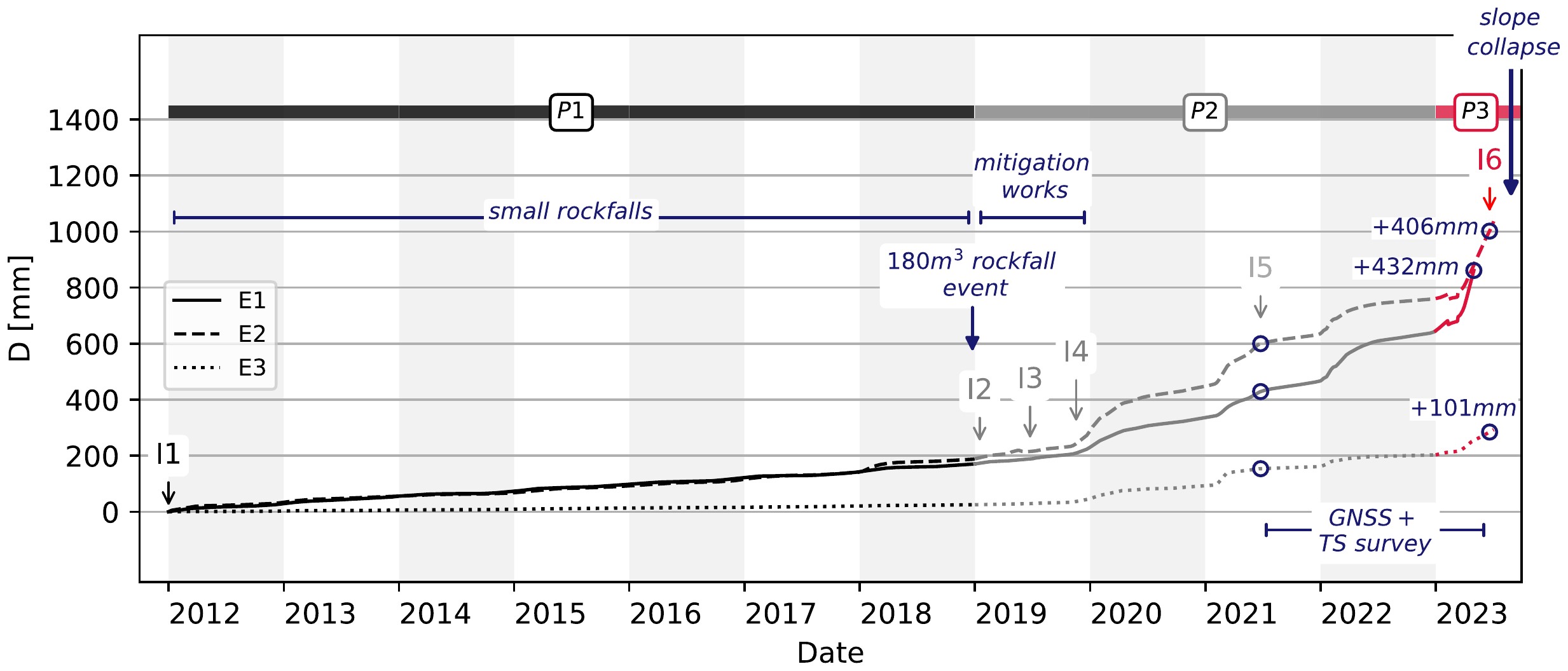}
\caption{Slope deformation time-series and rockfall activity at the La Praz slope. The figure focuses on the 2012--2023 period with concomitant deformation and seismic measurement campaigns. As a consequence, the 2006--2012 deformation data are not shown here. Extensometers E1 (continuous line) and E2 (dashed line) both measure the Main Rear Crack (MRC, Fig.~\ref{fig:site} and Fig.~\ref{fig:map}) opening between the stable slope S and compartment B2. Extensometer E3 (dotted line) measures the crack aperture between A1' and A1. The time series span across three slope displacement phases (SDP, see Section~\ref{sec:displ}): P1 (2012--2018, dark grey), P2 (2019--2022, grey) and P3 (2023, red). Measurements stopped a few weeks before the 27--28th August 2023 collapse because the extensometers reached their full scale. Passive seismic surveys I1 to I6 are shown with font colour according to SDP. Rockfall events and mitigation works are annotated with blue font. GNSS and total station (TS) survey period is shown in the bottom right corner. Concomitant extensometer aperture between I5 and I6 passive seismic surveys is shown with blue dots.}
\label{fig:defo}
\end{figure}

The progressive fracturing and subsidence of the La Praz unstable slope provokes frequent rockfalls, with unitary events scaling about 100--1000~m$^3$ reported during the 20th century and in 2002. Some of them impacted the covered railway and the local road located at the toe of the slope. After a relatively quiet period between 2006 and 2018, a 180~m$^3$ rockfall event occurred in 2018 (notably with a 6~m$^3$ block reaching the local road, Fig.~\ref{fig:defo}). Rock blasting and installation of hanging ring nets were performed in 2019 (see Fig.~\ref{fig:site}a and ``mitigation works'' in Fig.~\ref{fig:defo}). No major rockfall event was reported between 2019 and 2023. In late August 2023, the entire slope collapsed (estimated rockfall volume $\sim$13,000--15,000~m$^3$). In more details, a first 3000--5000~m$^3$ event occurred on the 27/08/2023 at 17\,h20 local time. Compartments A1 and A1' (Fig.~\ref{fig:site}b) detached and filled the embankment at the toe of the slope. A few blocks with large unitary size (about 30--40~m$^3$) reached the local road and the railway, interrupting all traffic. At least few hours later, i.e. during the night between 27 and 28/08/2023, the compartment A2 collapsed as well, followed by B1 and B2. The precise timing of this last sequence is poorly known because of the darkness. This second event scaled at about 10,000~m$^3$.

\section{Instrumentation}

A set of invar wire extensometers connected to a datalogger was set up across the main cracks in 2006. E1 and E2 extensometers were both set across the main rear crack while extensometer E3 spanned between compartments A1 and A1' (Fig.~\ref{fig:site}b and Fig.~\ref{fig:map}). No extensometric measurements were performed on compartment A2. As extensometers did not bring direct information about the external cliff motion, an automatic Leica\textsuperscript{\textregistered} MS60 total station was installed at the toe of the slope in 2018 (about 175~m in planar distance and 200~m below the unstable area). Total station targets are labelled Ti in Fig.~\ref{fig:site}b and Fig.~\ref{fig:map}. Both extensometer and total station operated with 1~h sampling. The three extensometers broke a few weeks before the collapse because of excessive aperture. We also conducted two GNSS measurement surveys with a Leica\textsuperscript{\textregistered} RTK GS07/GS08. Measurements had an accuracy of a few centimetres in the horizontal plane. Six characteristic points (Gi, Fig.~\ref{fig:site}b and Fig.~\ref{fig:map}) located at the top of the site were measured. These points could not be measured with the total station because of line-of-sight limitations. The GNSS campaigns were separated by two years: the first measurements were performed on the 25/06/2021 (concomitantly with passive seismic survey I5, cf. Table~\ref{tab:surveys}), the second on the 21/06/2023 (simultaneously with passive seismic survey I6), respectively.

Passive seismic measurements were performed on top of the La Praz unstable slope during successive surveys (see details in Table~\ref{tab:surveys}, including corresponding slope displacement phase SDP). The permanent continuous monitoring system I1 (2011--2013) was composed of one reference sensor on the adjacent stable slope S (I1--903, top left corner) and 2 sensors on the unstable slope (I1--901 on the right side of B2 compartment, and I1--902 on top of compartment A2 in Fig.~\ref{fig:map}). This instrument setup operated during SDP P1 (see details on SDP in Section~\ref{sec:displ}). The system was removed because of challenging power consumption in such remote, northern facing, $>$1000~m altitude alpine location and because of funding cutoff. Punctual surveys I2-I3-I4 were all three conducted in 2019 (I5 later in 2021) because of contractual reasons with the CD73 road operator. These surveys took place during SDP P2. The last passive seismic survey I6 (21/06/2023) was carried out only 2 months before the massive slope failure, during SDP P3. Unfortunately, no seismic measurement could be performed on the most prone-to-fall compartment A1 during survey I6 because of safety and accessibility issues. In the same line, no check measurements were performed on the stable massif during I6 for limited time reasons. For surveys I2-I6, the presence of the anchored ring net (Fig.~\ref{fig:site}b) reduced the accessibility of A1 and A1' for seismic instrumentation.

Air temperature was measured on site with a PT100 sensor and recorded every hour.

\begin{table}[htbp]
\centering
\caption{Description of successive passive seismic surveys at the La Praz site. Nb of M.P.: number of measurement points; R.L.: Record Length; $f_s$: sampling frequency. The background colour scales with the slope displacement phase (SDP) as defined in Fig.~\ref{fig:defo}.}
\label{tab:surveys}
\footnotesize
\setlength{\tabcolsep}{3.5pt}
\begin{tabular}{c>{\centering\arraybackslash}m{2.3cm}>{\centering\arraybackslash}m{2.0cm}c>{\centering\arraybackslash}m{2.3cm}>{\centering\arraybackslash}m{3.2cm}cc}
\toprule
Survey & Type of survey & Date(s) & SDP & Nb of M.P. & Seismic instruments & R.L. (min) & $f_s$ (Hz) \\
\midrule
I1 & Permanent, continuous passive seismic monitoring & 2011/04/29 -- 2013/07/15 & \Pone & 3 (2 sensors on the unstable site, 1 reference sensor) & Lennartz\textsuperscript{\textregistered} LE 3D 5s + Kinemetrics\textsuperscript{\textregistered} Granite & Continuous & 250 \\
\midrule
I2 & Punctual passive seismic survey & 2019/01/16 & \Ptwo & 4 & \multirow{5}{=}{\centering Mark Products\textsuperscript{\textregistered} (Sercel\textsuperscript{\textregistered}) L22-D 2.0~Hz + Leas\textsuperscript{\textregistered} (Keas) Cityshark II s/n 1880-0829-0001-PI} & 30 & 1\,000 \\
I3 & Punctual passive seismic survey & 2019/06/24 -- 2019/06/27 & \Ptwo & 21 & & 20 or 30 & 1\,000 \\
I4 & Punctual passive seismic survey & 2019/11/18 & \Ptwo & 4 & & 30 & 1\,000 \\
I5 & Punctual passive seismic survey & 2021/06/25 & \Ptwo & 5 & & 30 & 1\,000 \\
I6 & Punctual passive seismic survey & 2023/06/21 & \Pthree & 8 & & 20 & 1\,000 \\
\bottomrule
\end{tabular}
\end{table}

\section{Data processing}

We computed the 3D ground displacement of each point Gi between the two successive GNSS surveys (25/06/2021 and 21/06/2023). We selected the horizontal component of motion because it was predominant in the slope motion and well suited for map representation (Fig.~\ref{fig:map}). This horizontal displacement computed over 2 years was then converted into Mean Horizontal Ground Displacement Rate (MHGDR, expressed in mm/year). The same process was applied to total station measurements Ti over the same period.

The processing workflow for seismic data was based on obspy and scipy modules (Beyreuther et al., 2010; Krischer et al., 2015). We corrected the data from the instrument sensitivity and saved the seismic streams as miniseed files. We then computed the Horizontal to Vertical Spectral Ratio (HVSR) at each sensor's location following SESAME guidelines (Bard, 2005). HVSR were computed on 20--30~min long records (see Table~\ref{tab:surveys}) except for continuous survey I1 (1~h long records). We used 30~s tapered windows and performed the computation with the HtoV python toolbox (Krischer et al., 2015). The seismic sensors were leveled on the field using a bubble level. A polarization analysis (not shown here) revealed that HVSR amplification was maximal along a downwards plane striking N90\textdegree{} and dipping between 0\textdegree{} (i.e. purely horizontal plane) to 15\textdegree N (towards north). The mean dip was oriented 4.7\textdegree{} downward the slope, in very good agreement with previous observations in the literature (see introduction). We performed the rotation of components at each sensor before computing the HVSR: the North component was aligned with the dip angle while the Vertical component was set orthogonal to the dip plane. A Konno-Ohmachi filter with parameter $b = 90$ was used as a good trade-off between robustness and spectral peak resolution. The frequency of the main HVSR peak ---hereafter referred as $\fp$--- was automatically picked by a local maximum detection algorithm.

\section{Displacement monitoring}
\label{sec:displ}

\subsection{Spatial distribution of slope movements}

\begin{figure}[htbp]
\centering
\includegraphics[width=\textwidth]{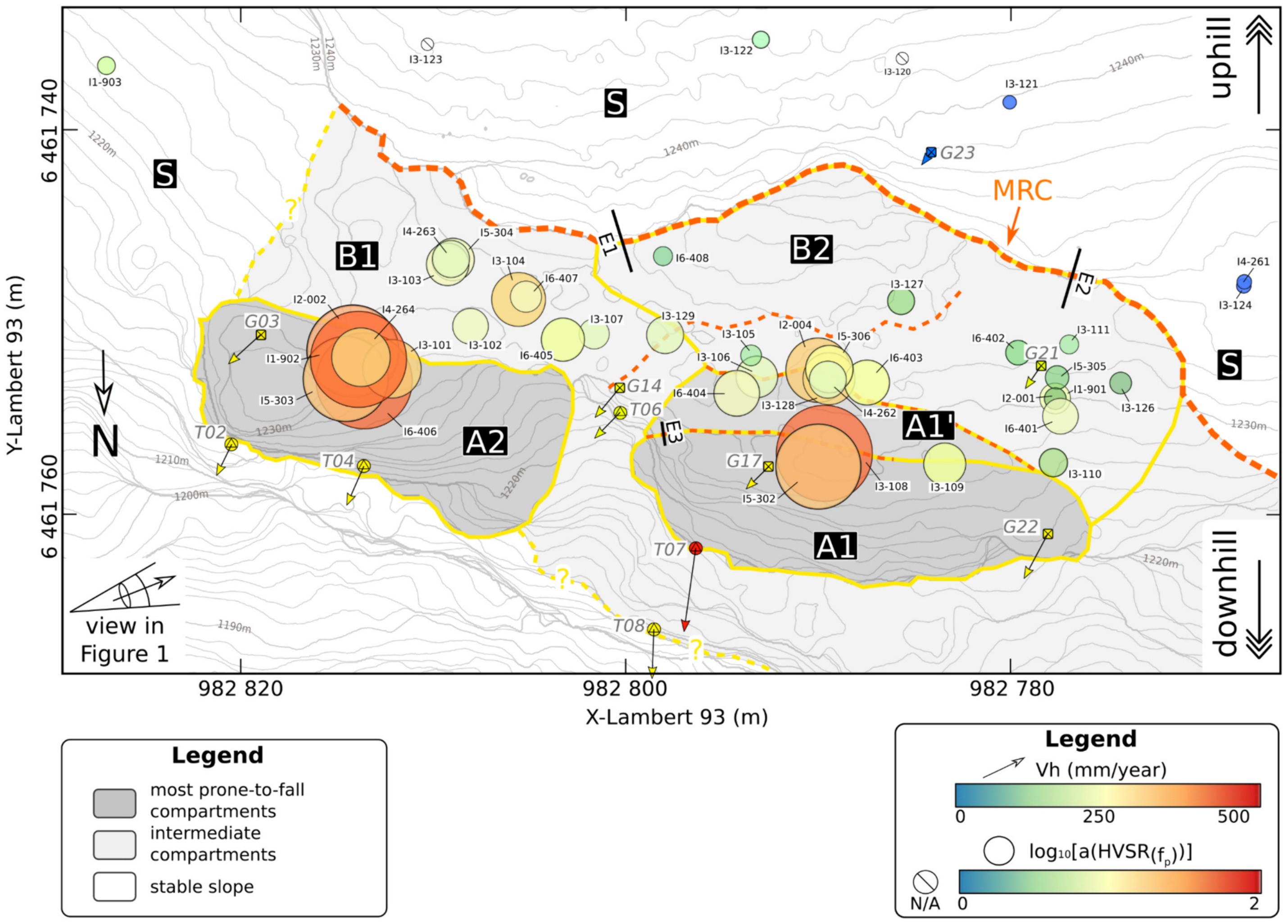}
\caption{Microzonation map of the site. The North direction (left side arrow) points downwards the figure. The map shows the angle of view used in Fig.~\ref{fig:site}a,b (bottom left corner) as well as uphill and downhill directions (top and bottom right corners, respectively). Open cracks are shown with dashed orange lines, including the main rear crack (MRC) bounding the site at south. The unstable compartments (yellow solid outline, A1-B2) and the adjacent stable slope (S) are labelled with black tags and highlighted with grey shading (see legend). Mean horizontal displacement rate for GNSS (Gi) and Leica total station (Ti) targets is shown with coloured arrows (see Vh colorbar and marker description in Fig.~\ref{fig:site}b). Extensometers E1, E2 and E3 are shown as black segments. Passive seismic results are shown with coloured dots. Both dot colour and dot size scale with the logarithm of HVSR main peak amplitude (see log$_{10}$[$\aH$] colorbar). See complete results in Table~\ref{tab:amp}. Sensor I3--125 (not visible here) is located outside of the map on the stable mass S.}
\label{fig:map}
\end{figure}

We focus here on the spatial heterogeneity of slope displacement at the La Praz site. The Mean Horizontal Ground Displacement Rate MHGDR is shown in Fig.~\ref{fig:site}b and Fig.~\ref{fig:map} for both GNSS (Gi) and total station (Ti) targets. The marker colour and the arrow length (Fig.~\ref{fig:map} only) both scale with velocities expressed in mm/year (see colorbar). The MHGDR appears very heterogeneous across the slope, ranging from a few millimeters to about 500~mm per year. The highest displacement velocities ($\sim$250--500~mm/year) are observed within the unstable slope, closest to the northern cliff face. Compartments A1 and A1' reveal a complex movement with large displacement measured at their top (G17 and G22 targets) but even greater lateral swelling detected at toe (T07 target, Fig.~\ref{fig:site}b and Fig.~\ref{fig:map}). This swelling is also visible below A1 toe (B1 area, T08 and T09 targets). The MHGDR then drops near the stable ledge (T12). The MHGDR for A2 compartment (G03 at top, T02 and T04 at toe) range about 200--300~mm/year. G14 and T06 targets located between A1' and A2 (compartment B1) show slightly smaller displacement rate than A2. MHGDR on top of B2 (target G21) is lower, ranging about a hundred of mm/year. This originates because B2 shows mainly subsidence behind A1 and A1' rather than horizontal displacement. The adjacent stable rock slope S shows a negligible displacement rate of a few mm/year (see check target G23) which lies within the measurement error range. We can note that the August 2023 collapse sequence described in Section~\ref{sec:rockfall} agreed very well with the MHGDR spatial distribution: compartments A1 and A1' showing the largest displacement rate falling first, followed by A2 a short time later.

\subsection{Slope deformation with time}

We use extensometer data to characterise the slope deformation with time because of long continuous operating time-series (see Fig.~\ref{fig:defo} focusing on the 2012--2023 period). We identified three successive slope displacement phases SDP (P1-P2-P3, Fig.~\ref{fig:defo}). During the SDP P1 (2012--2018 included), all three extensometers show progressive and linear fracture opening with time. The maximum opening rate scales about 30~mm/year for E1 and E2 set across the main rear crack (Fig.~\ref{fig:defo}). At the same time, compartment A1 separates from the adjacent A1' with rate of about 3~mm/year (E3, Fig.~\ref{fig:defo}). Starting with the 180~m$^3$ rockfall event, SDP P2 (2019--2022 included) showed a clear change in displacement rate with seasonal, successive bumps in fracture aperture. Slope deformation was maximal during winter and early spring. Fracture opening rate reached about 200~mm/year for E1 and E2, and $\sim$60~mm/year for E3. SDP P3 then started at the end of winter 2023 with a sharp increase in displacement rate and no slowdown during spring. Between I5 and I6 passive seismic surveys, the main rear crack opened of about +432, +406~mm (E1 and E2, Fig.~\ref{fig:defo}) while the crack between A1-A1' opened of +101~mm (E3). These values are consistent with the 2-year long total station and GNSS measurements that reported MHGDR of about a few hundreds of mm per year.

\section{HVSR peak origin}

\begin{figure}[htbp]
\centering
\includegraphics[width=\textwidth]{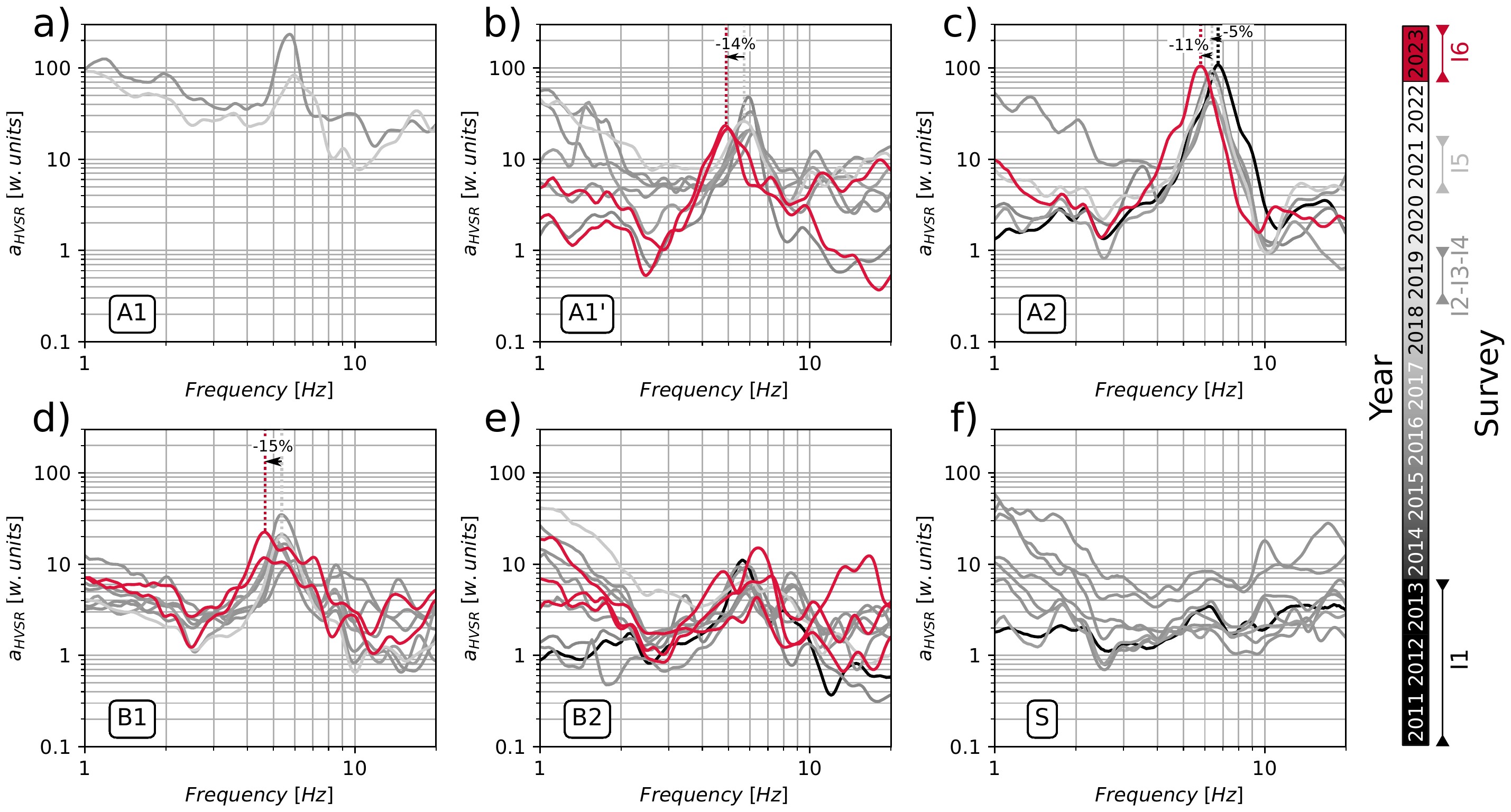}
\caption{Horizontal to Vertical Spectral Ratio (HVSR) curves observed on compartments A1 (a), A1' (b), A2 (c), B1 (d), B2 (e) and adjacent stable slope S (f). For each subplot, each curve corresponds to one measurement point (see list in Table~\ref{tab:surveys}). The curve colour scales with time (see colorbar). Corresponding survey is annotated on the right side of the colorbar. For I1 (2011--2013), HVSR was computed as the median over two years (black curves in subplots c, e, f). No measurements were conducted on A1 (subplot a) during survey I6 because of safety and accessibility issues. In the same line, no check measurements were conducted on the stable mass S (subplot f) during I6. Drops in $\fp$ are shown with arrows for A1', A2 and B1 (b,c,d, respectively).}
\label{fig:hvsr}
\end{figure}

All HVSR curves are shown in Fig.~\ref{fig:hvsr}, classified by rock compartment. Measurements located within the unstable slope (a-e) show clear HVSR peaks lying approximately in the range 3.0--10~Hz. These peaks are sharp, consistent and massive for A1, A1' and A2 (a, b, c, respectively). These peaks become smaller for B1 and B2 (d, e). The frequency of the peak (referred as $\fp$ in the rest of this article) lies generally between 5.5 and 7.0~Hz. Secondary HVSR peaks may be visible at higher frequency but are less consistent throughout the unstable rock slope. In contrast, the adjacent stable rock mass S (Fig.~\ref{fig:hvsr}f) exhibits rather smooth and flat ratios in the same frequency range, with plateau-like shape. The HVSR peaks observed in Fig.~\ref{fig:hvsr}a-e suggest a volume-controlled seismic resonance (Kleinbrod et al., 2019) occurring within the unstable, heavily fractured, softer rock body (Bottelin et al., 2013a).

\begin{figure}[htbp]
\centering
\includegraphics[width=\textwidth]{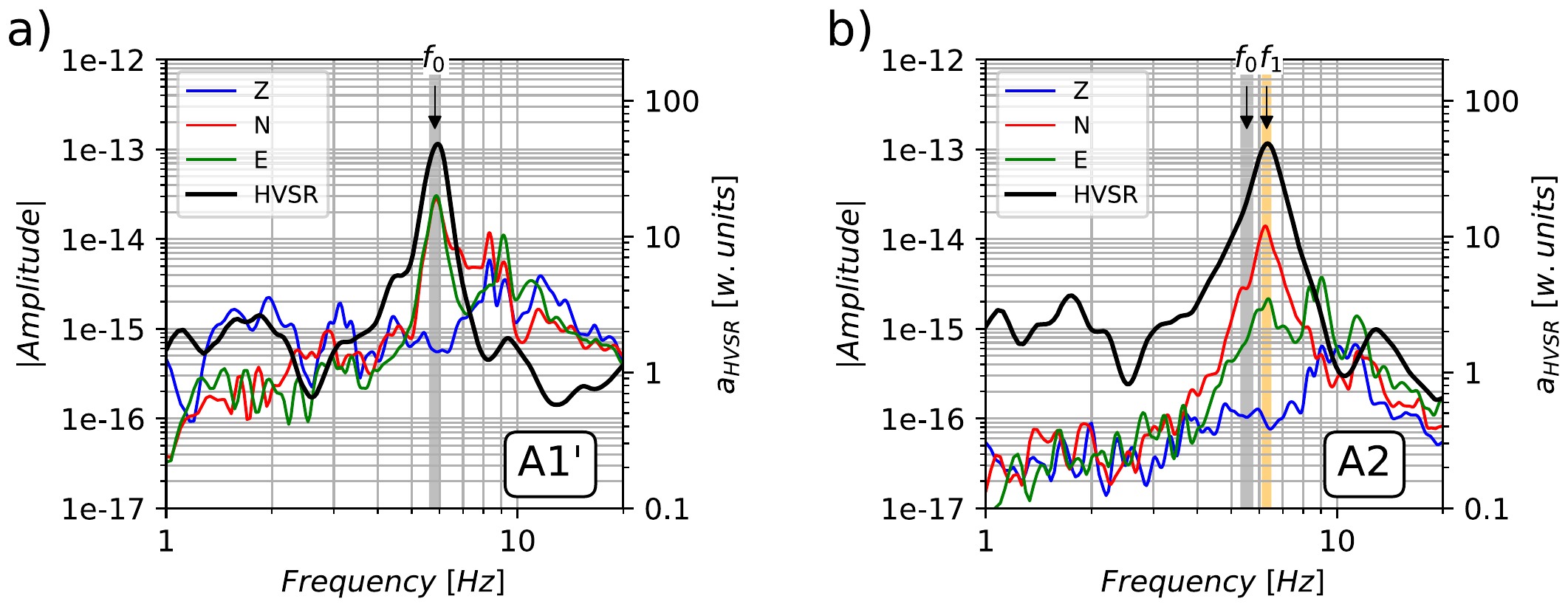}
\caption{Example of 3-component of motion spectra and HVSR curves a) for compartment A1' (measurement I2--004) and b) for compartment A2 (measurement I4--264). The spectra of Z, N and E components are shown with blue, red and green lines, respectively (left axis). The HVSR is shown with the thick black line (right axis).}
\label{fig:spectra}
\end{figure}

For most points, we observed that the HVSR peak $\fp$ originates from a sharp peak in North and East spectra (e.g. see A1' compartment in Fig.~\ref{fig:spectra}a). The frequency of the HVSR peak ($\fp$) hence corresponds exactly to the fundamental resonance frequency $f_0$ of the unstable slope (grey bar, Fig.~\ref{fig:spectra}a). In contrast, we observe a broader spectral peak for compartment A2 (e.g. I4--264 in Fig.~\ref{fig:spectra}b). The fundamental resonance frequency $f_0$ then appears only as a small lateral bump on the HVSR curve. A greater, adjacent peak (labelled $f_1$ in the rest of this article) is present at about 6.4~Hz. In this case only, the HVSR peak $\fp$ keeps track of compartment A2 specific resonance.

\section{Microzonation survey}
\label{sec:micro}

The amplitude of HVSR at peak frequency $\fp$ (hereafter referred to as $\aH$) is shown on a map in Fig.~\ref{fig:map}. The stable slope S shows either no peak (N/A) or peaks with very small amplitude ($\le$3.5). In contrast, points located within the unstable slope exhibit higher $\aH$ ranging from 4.2 to about 230 (see A1-A1'-A2-B1-B2 compartments in Fig.~\ref{fig:map}). This suggests a clear mechanical decoupling effect of the main rear crack. Compartments A1 and A2 located closest to the cliff show the greatest $\aH$ values (orange-red dots in Fig.~\ref{fig:map}). The compartments A1', B1 and especially B2 then show lower $\aH$ (orange, yellow and green markers, Fig.~\ref{fig:map}). The $\aH$ map suggests a strong mechanical role played by the fracture network. For example, a single $\sim$40~cm wide open crack between A1' and B2 causes a sharp drop in $\aH$ from 19 (I3-HVSR106) down to 4.7 (I3-HVSR105). This progressive reduction in $\aH$ across a fracture network, from the cliff face towards the stable rock mass has been observed and modelled in the literature (e.g. Burjánek et al., 2019; Häusler et al., 2021b, 2019; Kleinbrod et al., 2017a; Valentin et al., 2017).

\begin{figure}[htbp]
\centering
\includegraphics[width=0.5\textwidth]{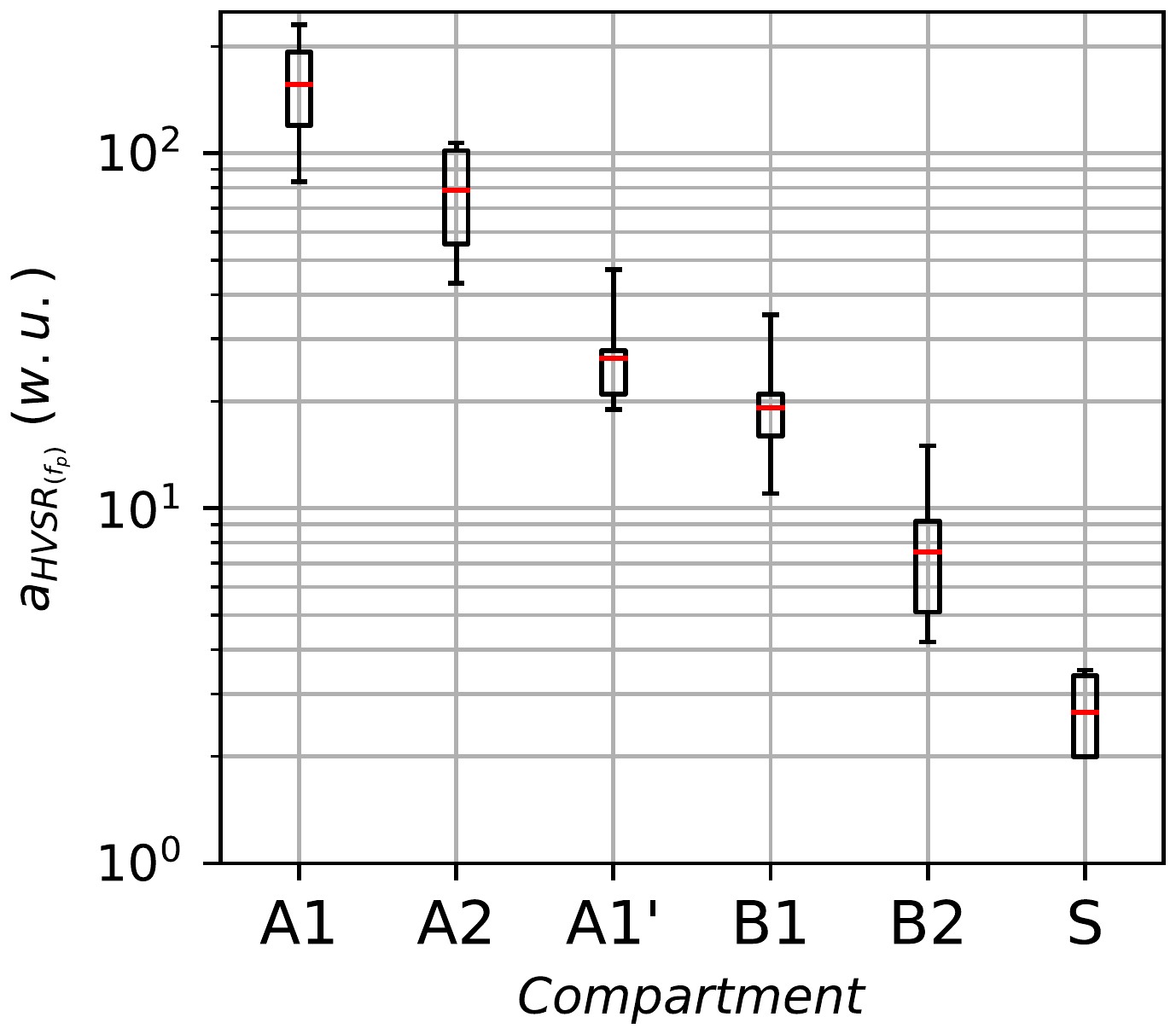}
\caption{Amplitude of HVSR main peak ($\aH$, y-axis) for compartments A1, A1', A2, B1, B2 and adjacent stable slope S (x-axis). For each group of data, the mean is shown with the red line. The box spans from the first quartile to third quartile. The whiskers cover the whole range of data, i.e. from the lowest to the greatest $\aH$ observed for each group. Number of data for each compartment: A1: 2 points; A2: 6 points; A1': 8 points; B1: 9 points; B2: 11 points; S: 8 points (see detail in Table~\ref{tab:amp}).}
\label{fig:box}
\end{figure}

Fig.~\ref{fig:box} summarises the $\aH$ statistics as a function of rock compartment (see full details in Table~\ref{tab:amp}). Compartments have been ordered with decreasing rockfall hazard (from the left to the right) as reported in geotechnical surveys (see Section~\ref{sec:site}). The mean $\aH$ (red line) varies over 2 orders of magnitude across the site. It shows a very good correlation with slope stability (the greater the $\aH$ the higher the rockfall hazard). This confirms that passive seismics can be used on complex unstable slope for mapping purposes, but also suggests possible quantitative hazard evaluation.

\begin{longtable}{>{\raggedright\arraybackslash}p{2.4cm}c>{\centering\arraybackslash}p{2.2cm}c>{\centering\arraybackslash}p{2.6cm}>{\centering\arraybackslash}p{2.6cm}}
\caption{Amplitude of HVSR peak at fundamental frequency ($\aH$) at the La Praz site. N/A: no clear peak. The slope displacement phase (SDP, fourth column) is recalled as background colour (see Fig.~\ref{fig:defo}).}\label{tab:amp}\\
\toprule
Location & Compartment & Measurement point & SDP & HVSR peak amplitude $\aH$ & Mean of HVSR peak amplitude $\aH$ \\
\midrule
\endfirsthead
\multicolumn{6}{l}{\textbf{Table \thetable} (continued)}\\
\toprule
Location & Compartment & Measurement point & SDP & HVSR peak amplitude $\aH$ & Mean of HVSR peak amplitude $\aH$ \\
\midrule
\endhead
\bottomrule
\endfoot
Unstable rock body & A1 & I3-108 & \Ptwo & 230 & 157 \\
 & & I5-302 & \Ptwo & 83 & \\
\cmidrule(l){2-6}
 & A2 & I1-902 & \Pone & 107 & 79 \\
 & & I2-002 & \Ptwo & 92 & \\
 & & I3-101 & \Ptwo & 43 & \\
 & & I4-264 & \Ptwo & 48 & \\
 & & I5-303 & \Ptwo & 78 & \\
 & & I6-406 & \Pthree & 105 & \\
\cmidrule(l){2-6}
 & A1' & I2-004 & \Ptwo & 47 & 26 \\
 & & I3-106 & \Ptwo & 19 & \\
 & & I3-109 & \Ptwo & 21 & \\
 & & I3-128 & \Ptwo & 33 & \\
 & & I4-262 & \Ptwo & 17 & \\
 & & I5-306 & \Ptwo & 26 & \\
 & & I6-403 & \Pthree & 24 & \\
 & & I6-404 & \Pthree & 24 & \\
\cmidrule(l){2-6}
 & B1 & I3-102 & \Ptwo & 16 & 19 \\
 & & I3-103 & \Ptwo & 21 & \\
 & & I3-104 & \Ptwo & 35 & \\
 & & I3-107 & \Ptwo & 11 & \\
 & & I3-129 & \Ptwo & 17 & \\
 & & I4-263 & \Ptwo & 18 & \\
 & & I5-304 & \Ptwo & 21 & \\
 & & I6-405 & \Pthree & 22 & \\
 & & I6-407 & \Pthree & 12 & \\
\cmidrule(l){2-6}
 & B2 & I1-901 & \Pone & 11 & 7.5 \\
 & & I2-001 & \Ptwo & 6 & \\
 & & I3-105 & \Ptwo & 4.7 & \\
 & & I3-110 & \Ptwo & 9.4 & \\
 & & I3-111 & \Ptwo & 4.2 & \\
 & & I3-126 & \Ptwo & 5.5 & \\
 & & I3-127 & \Ptwo & 9.0 & \\
 & & I5-305 & \Ptwo & 6.6 & \\
 & & I6-401 & \Pthree & 15.0 & \\
 & & I6-402 & \Pthree & 7.1 & \\
 & & I6-408 & \Pthree & 4.3 & \\
\midrule
Stable slope & S & I1-903 & \Pone & 3.5 & 2.7 \\
 & & I3-120 & \Ptwo & N/A & \\
 & & I3-121 & \Ptwo & 2.0 & \\
 & & I3-122 & \Ptwo & 3.5 & \\
 & & I3-123 & \Ptwo & N/A & \\
 & & I3-124 & \Ptwo & 2.0 & \\
 & & I3-125 & \Ptwo & 3.0 & \\
 & & I4-261 & \Ptwo & 2.0 & \\
\end{longtable}

\section{Temporal change detection}

In this section, we focus on the temporal changes in HVSR curves shown in Fig.~\ref{fig:hvsr} (see colorbar scaling with time and surveys). Only A2, B2 and S show measurements dating back from survey I1. HVSR curves for A2 (Fig.~\ref{fig:hvsr}c) suggest a slight progressive loss of about $-5$\% in $\fp$ between surveys I1 (black curve), I3 (grey) and later I5 (light grey). For the adjacent stable slope S (Fig.~\ref{fig:hvsr}f), the flat HVSR curve shape is very similar between surveys I1 to I2-I4. Variations in HVSR amplitude appear mostly due to sensor location during a single survey (i.e. dispersion across curves with the same colour) rather than changes with time. We do not observe any clear change for compartment B2 between I1 and I5 (Fig.~\ref{fig:hvsr}e). Most curves show a moderate HVSR peak (mean $\aH$ of 7.5) with some wandering in peak frequency $\fp$. For the unstable compartments A1-A1'-B1, no measurements were performed during I1 so that the comparison period is restricted to a narrower 2019--2021 period (I2-I5 surveys). We observe reproducible HVSR curve shape for A1' and B1 (Fig.~\ref{fig:hvsr}b and d), the main peak frequency $\fp$ yet showing slight wandering of about $\pm$0.5~Hz. We did not detect clear changes in HVSR amplitude between I2-I4 (2019, grey curves) and I5 (2021, light grey) on these A1' and B1 compartments. For A1 (Fig.~\ref{fig:hvsr}a), the curve shape varies between surveys I3 (2019, grey curve) and I5 (2021, light grey), the latter showing a broader but smaller peak and a slight rise in $\fp$. This observation on solely two measurements yet remained impossible to confirm during later survey I6 due to accessibility and safety issues on this compartment.

Contrasting with the relative HVSR curve stability described across I1-I5, the I6 survey performed in June 2023 (i.e. about 2 months prior to the massive slope collapse) exhibits clear and unequivocal changes in HVSR when compared to previous I5 survey (June 2021). The results are summarised in Table~\ref{tab:drops}. Substantial $-14$\%, $-11$\% and $-15$\% drops in HVSR peak frequency $\fp$ are detected for compartments A1', A2 and B1, respectively (see arrows in Fig.~\ref{fig:hvsr}b,c,d). These $\fp$ losses reflect a decrease in fundamental resonance frequency $f_0$ on both A1' and B1. For A2, $\fp$ loss mainly reflects a $f_1$ decrease (and possible $f_0$ drop recording as small lateral bump spotted at about 4.7~Hz, Fig.~\ref{fig:hvsr}c). No clear change pattern in HVSR curve shape nor peak frequency $\fp$ can be detected for B2 between I5 and I6 (Fig.~\ref{fig:hvsr}e). The only curve suggesting a decrease in $\fp$ on B2 comes from sensor I6--401 (HVSR peaking at 8 at about 4.9~Hz), which probably records adjacent compartment A1' vibration.

Whatever the compartment, we do not observe a clear, consistent trend in HVSR peak amplitude, $\aH$ remaining generally within the range of previous observed values. For example, it tends to remain stable for A1' but increases slightly for A2 and B1 (Fig.~\ref{fig:hvsr}b,c,d, respectively).

\begin{table}[htbp]
\centering
\caption{Drops in $\fp$ observed for the La Praz rock compartments between survey I5 (June 2021, SDP P2) and survey I6 (June 2023, SDP P3). Survey I6 occurred two months before the slope collapse. N/A: data not available. The background colour scales with the slope displacement phase (SDP) as defined in Fig.~\ref{fig:defo}.}
\label{tab:drops}
\small
\begin{tabular}{lcccc}
\toprule
Compartment & $\fp$ (Hz) -- I5 & $\fp$ (Hz) -- I6 & $\fp$ drop (Hz) & $\fp$ drop (\%) \\
\midrule
A1 & \cellcolor{sdpC}5.9 [I5-302] & \cellcolor{sdpD}\color{white}N/A & N/A & N/A \\
A1' & \cellcolor{sdpC}5.7 [I5-306] & \cellcolor{sdpD}\color{white}4.9 [I6-403 and I6-404] & $-0.8$ & $-14$ \\
A2 & \cellcolor{sdpC}6.4 [I5-303] & \cellcolor{sdpD}\color{white}5.7 [I6-406] & $-0.7$ & $-11$ \\
B1 & \cellcolor{sdpC}5.4 [I5-304] & \cellcolor{sdpD}\color{white}4.6 [I6-405 and I6-407] & $-0.8$ & $-15$ \\
B2 & \multicolumn{4}{c}{N/A} \\
S & \multicolumn{4}{c}{N/A} \\
\bottomrule
\end{tabular}
\end{table}

\section{Discussion}

In this article, we observed significant drops in HVSR peak frequency between surveys I5 (June 2021, SDP P2) and I6 (June 2023, SDP P3), the latter being performed 2 months before the slope collapse. The comparison of successive HVSR curves (Fig.~\ref{fig:hvsr}) revealed $-14$\%, $-11$\% and $-15$\% drops in $\fp$ affecting compartments A1', A2 and B1, respectively. Such real-case observation of resonance frequency drops on a complex, highly fractured unstable rock slope immediately prior to collapse is a first.

\begin{figure}[htbp]
\centering
\includegraphics[width=0.9\textwidth]{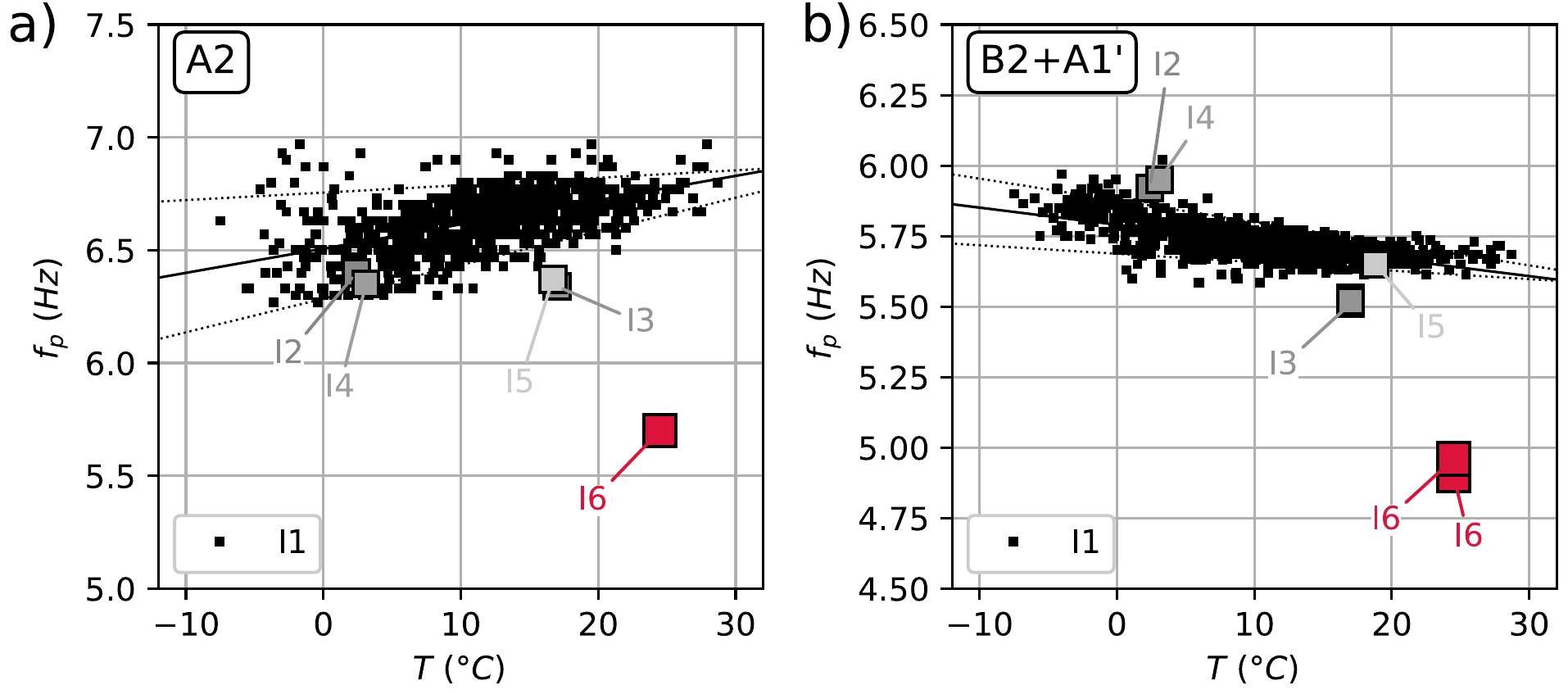}
\caption{a) HVSR peak frequency ($\fp$) of compartment A2 as a function of air temperature $T$. Measurements from survey I1 are shown as black squares, along with regression (solid black line) and 5\%--95\% quantiles (dotted black lines). Measurements from punctual surveys I2-I6 are shown with coloured squares (same colour scheme as in Table~\ref{tab:surveys} and Fig.~\ref{fig:hvsr}). b) Same as (a) for compartment B2 (survey I1, black squares) and then compartment A1' (surveys I2-I6).}
\label{fig:temp}
\end{figure}

In our opinion, these drops in $\fp$ are unambiguously caused by slope damage processes: indeed, the $\fp$ drop magnitude significantly exceeds the wandering due to environmental factors. We show $\fp$ tracking during I1-I6 as a function of air temperature $T$ for compartments A2 on one side (Fig.~\ref{fig:temp}a), and B2 + A1' together on the other (Fig.~\ref{fig:temp}b) respectively. As no measurements were conducted directly on A1' during survey I1, we selected a sensor located on B2 and showing the same resonance frequency at that time (I1--901). The ($T$, $\fp$) values during survey I1 (black squares) clearly show a positive (resp. negative) correlation with temperature for A2 (resp. B2). This originates probably from the domination of volume effects for B2 (dependence of Young's modulus to temperature) vs fracture effect domination for A2 (crack opening and closing via thermal expansion/contraction) (Bottelin et al., 2013b; Guillemot et al., 2022). $\fp$ tracking during punctual surveys I2-I5 are shown with coloured squares (same colour scheme as in Table~\ref{tab:surveys} and Fig.~\ref{fig:hvsr}). For both A2 and A1' compartments, ($T$, $\fp$) values lie relatively close to I1 data until I5 survey included (Fig.~\ref{fig:temp}a,b). We yet note that I3 and I5 measurements conducted during summer yield smaller $\fp$ values compared to I1. This is in good accordance with the slight $-5$\% drop in $\fp$ previously spotted for compartment A2 (Fig.~\ref{fig:hvsr}c). We yet recommend remaining conservative when interpreting small changes (a few percents only) in $\fp$ with punctual surveys. Punctual surveys are indeed more sensitive to anthropic perturbations and changes in ambient vibration wavefield because of shorter record duration. They could also suffer from instrumentation setup effects such as slight changes in sensor coupling with soil and/or levelling at each point (Bard, 2005). In contrast, on-site, permanent passive seismic instrumentation with longer recording duration offer better performance. Despite these experimental limitations regarding tiny changes in $\fp$, the considerable drops in $\fp$ ($-11$\% and $-14$\% for compartments A2 and A1', respectively) observed between surveys I5 and I6 are clearly visible on ($T$,$\fp$) plots. They are not explained by variations in temperature $T$. They clearly stand out from the environmental wandering observed over multiple years (I1 period, 2011--2013). This wandering is estimated to about 4.8\% for A2 and 3.9\% for A1' on the regression line (in the range $-5$ to 25\,\textdegree C, Fig.~\ref{fig:temp}a,b). These values lie well within the range of environmental-driven reversible changes reported in the literature, which range generally between 2 and 12\% at seasonal scale, see Colombero et al. (2021b) review and Geimer et al. (2022); Kleinbrod et al. (2019).

We also note that $\fp$ drops reported in this article are compatible with the slope deformation and the observed collapse sequence. The losses in $\fp$ indeed occurred between SDP P2 (2021--2022) and P3 (2023), notably characterised by intense slope deformation (e.g. additional +400~mm in E1 and E2, +150~mm in E3, Fig.~\ref{fig:defo}). In the same line, we highlight the fact that A1' showing a large drop in $\fp$ ($-14$\% between I5 and I6) was also the first to collapse ---along with A1--- on the 27/08/2023. We recall that unfortunately no passive seismic measurements could be carried out directly on A1 during I6 due to safety and accessibility issues. A few hours later, compartments A2 ($-11$\%) and B1 ($-15$\% in $\fp$) collapsed during the night between 27 and 28/08/2023. As B2 collapsed also on the same day without showing precursory $\fp$ decrease, we believe that its failure could be due to buttress removal (i.e. loss of A1-A1') rather than progressive weakening. We also observed a good agreement between the HVSR peak amplitude and slope collapse sequence. The three compartments A1, A1' and A2 which failed first indeed showed the largest $\aH$ of 157, 26 and 79, respectively (see Fig.~\ref{fig:box} and Table~\ref{tab:amp}). We note that although A1' shows smaller $\aH$ than A2, it fell together with A1: the latter acting probably as keystone. Interestingly, B2 (mean $\aH$ of 7.5) located behind A1' (mean $\aH$ of 26) did not fall concomitantly with A1-A1' on the 27/08/2023. In contrast, B2 remained attached longer to the stable rock mass, for at least a few hours (failure during the night 27--28/08/2023). This supports the fact that the fracture between A1' and B2 (clearly seen with $\aH$ mapping in Section~\ref{sec:micro}) acted as a major decoupling feature during the collapse sequence.

It is to be noted that i) our observation of $\fp$ drops related to slope damage and ii) the order of magnitude of these $\fp$ drops both fit well with the single previous case study, where a $-30$\% loss in $f_0$ was observed a few weeks before the rockfall (Chamousset site, Lévy et al., 2010). The La Praz (this study) and Chamousset sites yet showed very dissimilar geometry, failure mechanism, geology, orientation and elevation. Despite these disparities, $f_0$-$f_1$-$\fp$ drops ranging about ten to a few tens of percents tend to reflect significant mechanical loss in slope stability.

In this article, we presented very significant spatial variations in HVSR peak amplitude across the La Praz study site. We notably showed that $\aH$ can be used to rank the prone-to-fall degree of rock compartments and to delineate adjacent stable slope. We yet did not observe a clear rise in $\aH$ towards failure for a given rock compartment (Fig.~\ref{fig:hvsr}). For example, the $\aH$ observed just two months before slope collapse (survey I6) remained within the range of previously observed amplitudes even for A1' and A2 compartments (Fig.~\ref{fig:hvsr}b, c). This apparently contradicts with Fig.~\ref{fig:box} where $\aH$ scaled with the compartment prone-to-fall degree. It also contradicts numerical modelling suggesting a rise in $a_{HVSR}(f_0)$ with the progressive, massive rock bridge breaking between an unstable column and an adjacent stable slope (Valentin et al., 2017). We think that $\aH$ could be mainly controlled by the internal degree of rock fracturing, which was already high at the beginning of our experiment (survey I1) and did not change much since then. In contrast, $\fp$ could be more sensitive to boundary conditions (e.g. rock bridges breakage, basal sliding) as suggested by Valentin et al. (2017) for prone-to-fall rock columns.

\section{Conclusion and perspectives}

We showed a practical application of the HVSR technique to map the prone-to-fall degree of rock compartments. Our study dealt with an unstable, fractured slope with complex geometry and complex failure mechanism. The HVSR peak amplitude ($\aH$) varied from [26--157] on the most unstable zones, compared to [7.5--19] on intermediate areas and 2.7 on the adjacent stable rock mass. The HVSR mapping emphasized cracks that have the most mechanical decoupling importance. In a complementary manner, it hence may help delineating stable and safe areas within which no clear seismic resonance occurs.

We reported significant changes in passive seismic resonance of the La Praz site between 2021 and 2023, the last survey being carried out only two months before the massive slope collapse. Significant $-14$\%, $-11$\% and $-15$\% drops in HVSR peak frequency $\fp$ were observed on the three most prone-to-fall rock compartments A1', A2 and B1. These $\fp$ drops reflect losses in resonance frequencies related to preparatory rock material weakening. This case-study reporting such major irreversible changes in resonance frequencies for a complex and heavily fractured unstable slope is a first. It differs considerably from the single previous case study that dealt with a rather monolithic, slender limestone column. The La Praz site indeed shows a complex spatial resonance pattern and interlocking between rock sub-compartments. For such sites, global, in-depth information brought by ambient vibrations appear particularly valuable and complementary to surface displacement monitoring.

We think that future efforts should now focus on acquiring ambient vibration series on rock slopes until failure with a permanent instrumentation. Compared to the punctual successive surveys presented in this manuscript, a permanent monitoring should yield finer temporal details in terms of slope evolution. It could also reveal greater changes in seismic resonance occurring closer to slope failure, e.g. during the last two months in our case. This could be achieved with continuous 24/7 ambient vibration recording, yet at the cost of higher power consumption in a remote area. Another approach could consist in a triggered acquisition scheme (e.g. one or a few triggered records per day). Ambient vibration records should then follow the SESAME project guidelines (Bard, 2005): we would recommend at least 20--30~min record duration for a good estimate of the fundamental resonance peak (reasonably assuming that $\fp$ is $>$0.2--0.5~Hz). The record duration should be extended in case of strong nearby anthropic activity and/or weak resonance peak amplitude. The number and location of seismic sensors should be established during preliminary tests. For monolithic slender rock columns detaching from an adjacent rock mass, one sensor set up on top is generally sufficient to track the fundamental resonance peak. For complex rock slopes with several detaching, interlocked rock compartments (as shown in this article), a detailed microzonation survey could reveal necessary to design the passive seismic monitoring system and interpret the monitoring results.

\section*{CRediT authorship contribution statement}

\textbf{Pierre Bottelin:} Writing -- original draft, Validation, Supervision, Software, Methodology, Investigation, Formal analysis, Data curation. \textbf{Ombeline Méric:} Software, Investigation, Data curation. \textbf{Laurent Baillet:} Writing -- review \& editing, Validation, Investigation. \textbf{David Beniamine:} Software, Resources. \textbf{Anne Lescurier:} Project administration, Funding acquisition.

\section*{Declaration of competing interest}

The authors declare that they have no known competing financial interests or personal relationships that could have appeared to influence the work reported in this paper.

\section*{Data availability}

The data that has been used is confidential.

\section*{Acknowledgements}

The data from I1 survey were acquired thanks to the federative structure V.O.R. (Vulnérabilité des Ouvrages aux Risques) and the Alcotra project MASSA fundings. Seismic surveys I2-I6 were funded by the Conseil Départemental de la Savoie. The authors thank Denis Jongmans who initiated this work and for fruitful discussions. Thanks to Eric Larose for contributing to the development of the passive seismic technique on natural structures. Authors would like to thank L.~Lorier, J.~Bourdat and the SAGE instrumentation team for their support.

\section*{References}
\begingroup
\small
\setstretch{1.1}
\setlength{\parindent}{-1.5em}
\setlength{\leftskip}{1.5em}
\setlength{\parskip}{0.3em}

Arosio, D., Longoni, L., Papini, M., Scaioni, M., Zanzi, L., Alba, M., 2009. Towards rockfall forecasting through observing deformations and listening to microseismic emissions. Nat. Hazards Earth Syst. Sci. 9, 1119--1131. \url{https://doi.org/10.5194/nhess-9-1119-2009}.

Bard, P.Y., 2005. SESAME-Team (2005). Guidelines for the Implementation of the H/V Spectral Ratio Technique on Ambient Vibrations-Measurements, Processing and Interpretations. SESAME European research project.

Bessette-Kirton, E.K., Moore, J.R., Geimer, P.R., Finnegan, R., Häusler, M., Dzubay, A., 2022. Structural Characterization of a Toppling Rock Slab from Array-based Ambient Vibration Measurements and Numerical Modal Analysis. J. Geophys. Res. Earth 127, e2022JF006679. \url{https://doi.org/10.1029/2022JF006679}.

Beyreuther, M., Barsch, R., Krischer, L., Megies, T., Behr, Y., Wassermann, J., 2010. ObsPy: a Python toolbox for seismology. Seismol. Res. Lett. 81, 530--533. \url{https://doi.org/10.1785/gssrl.81.3.530}.

Bièvre, G., Franz, M., Larose, E., Carrière, S., Jongmans, D., Jaboyedoff, M., 2018. Influence of environmental parameters on the seismic velocity changes in a clayey mudflow (Pont-Bourquin Landslide, Switzerland). Engineering Geology 245, 248--257.

Bottelin, P., Jongmans, D., Baillet, L., Lebourg, T., Hantz, D., Lévy, C., Le Roux, O., Cadet, H., Lorier, L., Rouiller, J.-D., 2013a. Spectral analysis of prone-to-fall rock compartments using ambient vibrations. J. Environ. Eng. Geophys. 18, 205--217. \url{https://doi.org/10.2113/JEEG18.4.205}.

Bottelin, P., Lévy, C., Baillet, L., Jongmans, D., Gueguen, P., 2013b. Modal and thermal analysis of Les Arches unstable rock column (Vercors massif, French Alps). Geophys. J. Int. 194, 849--858. \url{https://doi.org/10.1093/gji/ggt046}.

Bottelin, P., Baillet, L., Larose, E., Jongmans, D., Hantz, D., Brenguier, O., Cadet, H., Helmstetter, A., 2017. Monitoring rock reinforcement works with ambient vibrations: La Bourne case study (Vercors, France). Eng. Geol. 226, 136--145. \url{https://doi.org/10.1016/j.enggeo.2017.06.002}.

Bottelin, P., Baillet, L., Carrier, A., Larose, E., Jongmans, D., Brenguier, O., Cadet, H., 2021. Toward Workable and Cost-Efficient monitoring of Unstable Rock Compartments with Ambient Noise. Geosciences 11, 242. \url{https://doi.org/10.3390/geosciences11060242}.

Brincker, R., Ventura, C., 2015. Introduction to Operational Modal Analysis. John Wiley \& Sons.

Burjánek, J., Gassner-Stamm, G., Poggi, V., Moore, J.R., Fäh, D., 2010. Ambient vibration analysis of an unstable mountain slope. Geophys. J. Int. 180, 820--828. \url{https://doi.org/10.1111/j.1365-246X.2009.04451.x}.

Burjánek, J., Moore, J.R., Yugsi Molina, F.X., Fäh, D., 2012. Instrumental evidence of normal mode rock slope vibration. Geophys. J. Int. 188, 559--569. \url{https://doi.org/10.1111/j.1365-246X.2011.05272.x}.

Burjánek, J., Edwards, B., Fäh, D., 2014. Empirical evidence of local seismic effects at sites with pronounced topography: a systematic approach. Geophys. J. Int. 197, 608--619.

Burjánek, J., Gischig, V., Moore, J.R., Fäh, D., 2018. Ambient vibration characterization and monitoring of a rock slope close to collapse. Geophys. J. Int. 212, 297--310. \url{https://doi.org/10.1093/gji/ggx424}.

Burjánek, J., Kleinbrod, U., Fäh, D., 2019. Modeling the Seismic Response of Unstable Rock Mass with Deep Compliant Fractures. J. Geophys. Res. Solid Earth 124, 13039--13059. \url{https://doi.org/10.1029/2019JB018607}.

Colombero, C., Baillet, L., Comina, C., Jongmans, D., Vinciguerra, S., 2017. Characterization of the 3-D fracture setting of an unstable rock mass: from surface and seismic investigations to numerical modeling: Fracture Settings of an Unstable Cliff. J. Geophys. Res. Solid Earth 122, 6346--6366. \url{https://doi.org/10.1002/2017JB014111}.

Colombero, C., Baillet, L., Comina, C., Jongmans, D., Larose, E., Valentin, J., Vinciguerra, S., 2018. Integration of ambient seismic noise monitoring, displacement and meteorological measurements to infer the temperature-controlled long-term evolution of a complex prone-to-fall cliff. Geophys. J. Int. 213, 1876--1897. \url{https://doi.org/10.1093/gji/ggy090}.

Colombero, C., Godio, A., Jongmans, D., 2021a. Ambient Seismic Noise and Microseismicity monitoring of a Prone-To-fall Quartzite Tower (Ormea, NW Italy). Remote Sens. (Basel) 13, 1664. \url{https://doi.org/10.3390/rs13091664}.

Colombero, C., Jongmans, D., Fiolleau, S., Valentin, J., Baillet, L., Bièvre, G., 2021b. Seismic Noise Parameters as Indicators of Reversible modifications in Slope Stability: a Review. Surv. Geophys. 42, 339--375. \url{https://doi.org/10.1007/s10712-021-09632-w}.

Debelmas, J., 1989. Carte géologique de la France à 1:50 000 Feuille 775 -- Modane (Document cartographique), Carte géologique de la France à 1:50 000. Bureau de recherches géologiques et minières, Orléans.

Farvacque, M., Lopez-Saez, J., Corona, C., Toe, D., Bourrier, F., Eckert, N., 2019. How is rockfall risk impacted by land-use and land-cover changes? Insights from the French Alps. Global Planet. Change 174, 138--152. \url{https://doi.org/10.1016/j.gloplacha.2019.01.009}.

Finnegan, R., Moore, J.R., Geimer, P.R., 2021. Vibration of natural rock arches and towers excited by helicopter-sourced infrasound. Earth Surf. Dyn. 9, 1459--1479. \url{https://doi.org/10.5194/esurf-9-1459-2021}.

Finnegan, R., Moore, J.R., Geimer, P.R., Dzubay, A., Bessette-Kirton, E.K., Bodtker, J., Vollinger, K., 2022. Ambient Vibration Modal Analysis of Natural Rock Towers and Fins. Seismol. Res. Lett. 93, 1777--1786. \url{https://doi.org/10.1785/0220210325}.

Fu, Z.-F., He, J., 2001. Modal analysis. Elsevier. \url{https://doi.org/10.1016/B978-0-7506-5079-3.X5000-1}.

Geimer, P.R., Finnegan, R., Moore, J.R., 2020. Sparse ambient resonance measurements reveal dynamic properties of freestanding rock arches. Geophys. Res. Lett. 47, e2020GL087239. \url{https://doi.org/10.1029/2020GL087239}.

Geimer, P.R., Finnegan, R., Moore, J.R., 2022. Meteorological Controls on Reversible Resonance changes in Natural Rock Arches. J. Geophys. Res. Earth 127, e2022JF006734. \url{https://doi.org/10.1029/2022JF006734}.

Gschwind, S., Loew, S., Wolter, A., 2019. Multi-stage structural and kinematic analysis of a retrogressive rock slope instability complex (Preonzo, Switzerland). Engineering Geology 252, 27--42. \url{https://doi.org/10.1016/j.enggeo.2019.02.018}.

Guillemot, A., Baillet, L., Larose, E., Bottelin, P., 2022. Changes in resonance frequency of rock columns due to thermoelastic effects on a daily scale: observations, modelling and insights to improve monitoring systems. Geophys. J. Int. 231, 894--906. \url{https://doi.org/10.1093/gji/ggac216}.

Häusler, M., Michel, C., Burjánek, J., Fäh, D., 2019. Fracture Network Imaging on Rock Slope Instabilities using Resonance Mode Analysis. Geophys. Res. Lett. 46, 6497--6506. \url{https://doi.org/10.1029/2019GL083201}.

Häusler, M., Geimer, P.R., Finnegan, R., Fäh, D., Moore, J.R., 2021a. An update on techniques to assess normal-mode behavior of rock arches by ambient vibrations. Earth Surf. Dyn. 9, 1441--1457. \url{https://doi.org/10.5194/esurf-9-1441-2021}.

Häusler, M., Michel, C., Burjánek, J., Fäh, D., 2021b. Monitoring the Preonzo Rock Slope Instability using Resonance Mode Analysis. J. Geophys. Res. Earth 126, e2020JF005709. \url{https://doi.org/10.1029/2020JF005709}.

Häusler, M., Gischig, V., Thöny, R., Glueer, F., Donat, F., 2022. Monitoring the changing seismic site response of a fast-moving rockslide (Brienz/Brinzauls, Switzerland). Geophys. J. Int. 229, 299--310. \url{https://doi.org/10.1093/gji/ggab473}.

Jaboyedoff, M., Del Gaudio, V., Derron, M.-H., Grandjean, G., Jongmans, D., 2019. Characterizing and monitoring landslide processes using remote sensing and geophysics. Engineering Geology 259, 105167. \url{https://doi.org/10.1016/j.enggeo.2019.105167}.

Kleinbrod, U., Burjánek, J., Fäh, D., 2017a. On the seismic response of instable rock slopes based on ambient vibration recordings. Earth Planets Space 69, 126. \url{https://doi.org/10.1186/s40623-017-0712-5}.

Kleinbrod, U., Burjánek, J., Hugentobler, M., Amann, F., Fäh, D., 2017b. A comparative study on seismic response of two unstable rock slopes within same tectonic setting but different activity level. Geophys. J. Int. 211, 1428--1448. \url{https://doi.org/10.1093/gji/ggx376}.

Kleinbrod, U., Burjánek, J., Fäh, D., 2019. Ambient vibration classification of unstable rock slopes: a systematic approach. Engineering Geology 249, 198--217. \url{https://doi.org/10.1016/j.enggeo.2018.12.012}.

Krischer, L., Megies, T., Barsch, R., Beyreuther, M., Lecocq, T., Caudron, C., Wassermann, J., 2015. ObsPy: a bridge for seismology into the scientific Python ecosystem. Computational Science \& Discovery 8, 014003. \url{https://doi.org/10.1088/1749-4699/8/1/014003}.

Larose, E., Carrière, S., Voisin, C., Bottelin, P., Baillet, L., Guéguen, P., Walter, F., Jongmans, D., Guillier, B., Garambois, S., 2015. Environmental seismology: what can we learn on earth surface processes with ambient noise? J. Appl. Geophys. 116, 62--74. \url{https://doi.org/10.1016/j.jappgeo.2015.02.001}.

Le Breton, M., Bontemps, N., Guillemot, A., Baillet, L., Larose, É., 2021. Landslide monitoring using seismic ambient noise correlation: challenges and applications. Earth Sci. Rev. 216, 103518. \url{https://doi.org/10.1016/j.earscirev.2021.103518}.

Lévy, C., Baillet, L., Jongmans, D., Mourot, P., Hantz, D., 2010. Dynamic response of the Chamousset rock column (Western Alps, France). J. Geophys. Res. Earth 115. \url{https://doi.org/10.1029/2009JF001606}.

Lotti, A., Pazzi, V., Saccorotti, G., Fiaschi, A., Matassoni, L., Gigli, G., 2018. HVSR Analysis of Rockslide Seismic Signals to Assess the Subsoil Conditions and the Site Seismic Response. International Journal of Geophysics 2018, e9383189. \url{https://doi.org/10.1155/2018/9383189}.

Maheshwari, S., Bhowmik, R., Samanta, M., 2023. Rockfall Hazard: A Comprehensive Review of Current Mitigation Practices. In: Thambidurai, P., Singh, T.N. (Eds.), Landslides: Detection, Prediction and Monitoring: Technological Developments. Springer International Publishing, Cham, pp. 175--209. \url{https://doi.org/10.1007/978-3-031-23859-8_9}.

Moore, J.R., Gischig, V., Burjánek, J., Loew, S., Fäh, D., 2011. Site Effects in Unstable Rock Slopes: Dynamic Behavior of the Randa Instability (Switzerland). Bull. Seismol. Soc. Am. 101, 3110--3116. \url{https://doi.org/10.1785/0120110127}.

Moore, J.R., Thorne, M.S., Koper, K.D., Wood, J.R., Goddard, K., Burlacu, R., Doyle, S., Stanfield, E., White, B., 2016. Anthropogenic sources stimulate resonance of a natural rock bridge. Geophys. Res. Lett. 43, 9669--9676. \url{https://doi.org/10.1002/2016GL070088}.

Moore, J.R., Geimer, P.R., Finnegan, R., Michel, C., 2019. Dynamic Analysis of a large Freestanding Rock Tower (Castleton Tower, Utah). Bull. Seismol. Soc. Am. 109, 2125--2131. \url{https://doi.org/10.1785/0120190118}.

Müller, J., Burjánek, J., 2023. In situ estimation of effective rock elastic moduli by seismic ambient vibrations. Int. J. Rock Mech. Min. Sci. 170, 105459. \url{https://doi.org/10.1016/j.ijrmms.2023.105459}.

Rose, N.D., Hungr, O., 2007. Forecasting potential rock slope failure in open pit mines using the inverse-velocity method. Int. J. Rock Mech. Min. Sci. 44, 308--320. \url{https://doi.org/10.1016/j.ijrmms.2006.07.014}.

Rosser, N., Lim, M., Petley, D., Dunning, S., Allison, R., 2007. Patterns of precursory rockfall prior to slope failure. J. Geophys. Res. 112, F04014. \url{https://doi.org/10.1029/2006JF000642}.

Starr, A.M., Moore, J.R., Thorne, M.S., 2015. Ambient resonance of Mesa Arch, Canyonlands National Park, Utah. Geophys. Res. Lett. 42, 6696--6702. \url{https://doi.org/10.1002/2015GL064917}.

Valentin, J., Capron, A., Jongmans, D., Baillet, L., Bottelin, P., Donze, F., Larose, E., Mangeney, A., 2017. The dynamic response of prone-to-fall columns to ambient vibrations: comparison between measurements and numerical modelling. Geophys. J. Int. 208, 1058--1076. \url{https://doi.org/10.1093/gji/ggw440}.

Xie, F., Larose, E., Wang, Q., Zhang, Y., 2023. In-situ monitoring of rock slope destabilization with ambient seismic noise interferometry in Southwest China. Engineering Geology 312, 106922. \url{https://doi.org/10.1016/j.enggeo.2022.106922}.

\endgroup

\end{document}